\documentclass[12pt]{article}
\usepackage{amsmath,amssymb}
\usepackage{epsfig}

\newcommand{\ep}{\varepsilon}

\catcode`@=11
\newcount\@tempcntc
\def\@citex[#1]#2{\if@filesw\immediate\write\@auxout{\string\citation{#2}}\fi
  \@tempcnta\z@\@tempcntb\m@ne\def\@citea{}\@cite{\@for\@citeb:=#2\do
    {\@ifundefined
       {b@\@citeb}{\@citeo\@tempcntb\m@ne\@citea\def\@citea{,}{\bf ?}\@warning
       {Citation `\@citeb' on page \thepage \space undefined}}%
    {\setbox\z@\hbox{\global\@tempcntc0\csname b@\@citeb\endcsname\relax}%
     \ifnum\@tempcntc=\z@ \@citeo\@tempcntb\m@ne
       \@citea\def\@citea{,}\hbox{\csname b@\@citeb\endcsname}%
     \else
      \advance\@tempcntb\@ne
      \ifnum\@tempcntb=\@tempcntc
      \else\advance\@tempcntb\m@ne\@citeo
      \@tempcnta\@tempcntc\@tempcntb\@tempcntc\fi\fi}}\@citeo}{#1}}
\def\@citeo{\ifnum\@tempcnta>\@tempcntb\else\@citea\def\@citea{,}%
  \ifnum\@tempcnta=\@tempcntb\the\@tempcnta\else
   {\advance\@tempcnta\@ne\ifnum\@tempcnta=\@tempcntb \else \def\@citea{--}\fi
    \advance\@tempcnta\m@ne\the\@tempcnta\@citea\the\@tempcntb}\fi\fi}
\catcode`@=12
\begin{document}

%
%

\title{
\vskip-3cm{\baselineskip14pt
\centerline{\normalsize DESY~12--068\hfill ISSN~0418--9833}
\centerline{\normalsize May 2012\hfill}}
\vskip1.5cm
Mellin-Barnes representations of Feynman diagrams, linear systems of
differential equations, and polynomial solutions
}

\author{
{\sc Mikhail~Yu.~Kalmykov}\thanks{On leave of absence from
Joint Institute for Nuclear Research, 141980 Dubna (Moscow Region), Russia.},
{\sc Bernd~A.~Kniehl}
\\
\\
{\normalsize II. Institut f\"ur Theoretische Physik, Universit\"at Hamburg,}\\
{\normalsize Luruper Chaussee 149, 22761 Hamburg, Germany}
}

\date{}

\maketitle
\abstract{
We argue that the Mellin-Barnes representations of Feynman diagrams can be used
for obtaining linear systems of homogeneous differential equations for the
original Feynman diagrams with arbitrary powers of propagators without recourse
to the integration-by-parts technique. 
These systems of differential equations can be used
(i) for the differential reductions to sets of basic functions and 
(ii) for counting the numbers of master integrals. 
\medskip

\noindent
PACS numbers: 02.30.Gp, 02.30.Lt, 11.15.Bt, 12.38.Bx\\
Keywords: 
Feynman diagrams, 
Hypergeometric functions,
Mellin-Barnes representations
}

\newpage

\renewcommand{\thefootnote}{\arabic{footnote}}
\setcounter{footnote}{0}

%

\renewcommand{\thefootnote}{\arabic{footnote}}
\setcounter{footnote}{0}

\section{Introduction}

The importance of the differential-equation approach to the description of the
analytical properties of Feynman diagrams has been recognized a long time ago
\cite{Regge}.
Within dimension regularization \cite{dimreg}, the differential-equation
technique \cite{DE}, based on the integration-by-part (IBP) relations
\cite{ibp}, has been one of the most popular tools for the analytical
evaluation of Feynman diagrams during the last decade \cite{review}. 
Here we argue that linear systems of homogeneous differential equations may be
derived for Feynman diagrams starting from their Mellin-Barnes representations
without resorting to IBP relations \cite{idea}.

Our staring point is the multiple Mellin-Barnes representation of Feynman
diagrams \cite{calan,BD,smirnov}, which may be written in the following form: 
\begin{eqnarray}
\Phi({\bf A},\vec{B};{\bf C}, \vec{D};\vec{z}) 
& = & 
\int_{-i \infty}^{+i \infty}
d\vec{t}
\phi(\vec{t}) 
\vec{z}^{\,\vec{t}} 
\nonumber \\  
& = & 
\mbox{Const}
\int_{-i \infty}^{+i \infty}
\prod_{a,b,c,r}
dt_c
\frac{\Gamma(\sum_{i=1}^m A_{ai}t_i \!+\! B_a)}{\Gamma(\sum_{j=1}^r C_{bj}t_j \!+\! D_{b})}
 z_k^{\sum_l \alpha_{kl} t_l}
\;,
\label{MB}
\end{eqnarray}
where $z_k$ are ratios of Mandelstam variables and
${\bf A},\vec{B},{\bf C},\vec{D},\mbox{\boldmath$\alpha$}$ are matrices
and vectors depending linearly on the dimension $n$ of space-time and the
powers of the propagators. 
An important property of Feynman diagrams is that the matrices ${\bf A}$ and
${\bf C}$ only include integers.
Let us define the polynomials $P_i$ and $Q_i$ as 
\begin{equation}
\frac{P_i(\vec{t})}{Q_i(\vec{t})} = \frac{\phi(\vec{t}+\vec{e_i})}{\phi(\vec{t})} \;,
\end{equation}
where $\vec{e_i}$ is a unit vector with nonzero element at the $i$-th place.
Then, the integral in Eq.~(\ref{MB}) satisfies the following linear system of
homogeneous differential equations: 
\begin{eqnarray}
\left. Q_i(\vec{t}) \right|_{t_j \to \theta_j }\frac{1}{z_i} \Phi({\bf A},\vec{B};{\bf C}, \vec{D};\vec{z})  
= \left. P_i(\vec{t}) \right|_{t_j \to \theta_j } \Phi({\bf A},\vec{B};{\bf C}, \vec{D};\vec{z})  \;,
\label{MB:DE}
\end{eqnarray}
where $\theta_i =  z_id/dz_i$, we assume that, in Eq.~(\ref{MB}), the number of
variables $z_i$ is equal to number of integration variables $t_i$, and, for
simplicity, we put $\alpha_{ij}=\delta_{ij}$.
Moreover, the function $\Phi$ defined in Eq.~(\ref{MB}) satisfies the
differential contiguous relations: 
\begin{eqnarray}
\Phi({\bf A}, \vec{B} \!+\! \vec{e_a}; {\bf C}, \vec{D};\vec{z}) 
& = & 
\left(\sum_{i=1}^m A_{ai} \theta_i \!+\! B_a \right) 
\Phi({\bf A},\vec{B}; {\bf C}, \vec{D};\vec{z}) \;,
\nonumber \\  
\quad 
\Phi({\bf A}, \vec{B}; {\bf C}, \vec{D} \!-\! \vec{e_b};\vec{z}) 
& = &  
\left(\sum_{j=1}^r C_{bj} \theta_j \!+\! D_{b} \right)
\Phi({\bf A},\vec{B}; {\bf C}, \vec{D};\vec{z}) 
\;.
\label{step-up-down}
\end{eqnarray}
This system of differential equations may be analyzed using the Gr\"obner basis
technique  \cite{Grobner}.
Specifically, 
(i) the holonomic rank $r$ of the system in Eq.~(\ref{MB:DE}) can be
evaluated,\footnote{%
This can also be done using a finite number of prolongations and converting
the original system into a Pfaffian system \cite{cartan}.} and
(ii) starting from the direct differential operators in
Eq.~(\ref{step-up-down}) and the system of differential equations in
Eq.~(\ref{MB:DE}), the inverse differential operators can be constructed
\cite{theorem}.
By the action of such differential operators on the function $\Phi$, the value
of any parameter $B_j, D_k$ may be shifted by an arbitrary integer.
This algorithm is working for holonomic functions if the number of linearly
independent solutions is finite. 
As was shown in Ref.~\cite{KK}, within analytical regularization \cite{Speer}, 
Feynman diagrams satisfy holonomic systems of linear differential equations
under the condition that all particles have different masses.
In fact, this statement is the basis of the algorithm proposed in
Ref.~\cite{Tarasov:1998}.
To our understanding, a rigorous proof for the case of on-shell diagrams or
diagrams with zero internal masses does not yet exist, albeit very interesting
work has been done in this direction \cite{D}.
For our analysis, it is sufficient to assume that there is a set of independent
variables $z_i$ for each Mellin-Barnes integral\footnote{ 
The application of the summation technique to Mellin-Barnes integrals with
$z_i=1$ was discussed in Ref.~\cite{Blumlein}.}
in Eq.~(\ref{MB}). 
Moreover, it has been shown \cite{beukers} recently that Mellin-Barnes
integrals satisfy systems of differential equations corresponding to
Gelfand-Kapranov-Zelevisky hypergeometric equations \cite{gelfand}. 
Another necessary condition is that two contours differing by a translation by
one unit along the real axis are equivalent. 
From the analysis performed in Refs.~\cite{BD,resolution}, we expect that this
statement is valid for all Feynman diagrams before constructing
$\ep$ expansions. 

The aim of the present paper is to illustrate this approach and to outline
how it can be used to count master integrals, considering as examples one-fold
Mellin-Barnes integrals and the corresponding Feynman diagrams.
In fact, the evaluation of master integrals is often the bottleneck of
multi-loop calculations of both Feynman diagrams and scattering amplitudes,
especially if several different mass scales are involved, and any opportunity
to reduce their number below the achievements of the standard techniques of
solving IBP relations, as implemented in various publicly available computer
codes, should be highly welcome.
The technique advocated here may also allow one to gain deeper insights into
the mathematical structures of multi-scale Feynman diagrams.

\section{One-fold Mellin-Barnes integral} 

Let us consider the function 
\begin{eqnarray}
\Phi(\vec{A},\vec{B};\vec{C}, \vec{D};z;r)
& = &  
C_\Phi
\int_{\gamma+iR} dt
\frac{
\prod_{i=0}^{K} \Gamma\left(A_i+t\right)
\prod_{j=0}^{L} \Gamma\left(C_j-t\right)
}
{
\prod_{k=0}^{R} \Gamma\left(B_k + t\right)
\prod_{l=0}^{J} \Gamma\left(D_l - t\right)}
\Gamma(-t) 
z^{t+r} 
\; , 
\label{Phi1}
\end{eqnarray}
where $C_\Phi$ is some $z$-independent constant depending on ratios of $\Gamma$
functions with arguments being linear combinations of powers of propagators and
the space-time dimension $n$. 
In the remainder of this letter, we assume that $r=0$.
In order to restore a nonzero value of $r$, it is sufficient to substitute
$\theta \to \theta - r$.
Let us assume that the differences between any two parameters $A_i,B_j,C_k,D_l$
are not integers. 
Then, this function satisfies the following homogeneous differential equation:
\begin{eqnarray}
&&
(-1)^{L+1} 
\theta
\prod_{i=1}^R (\theta \!+\! B_i \!-\! 1)
\prod_{j=1}^L (\theta \!-\! C_j)
\Phi(\vec{A},\vec{B};\vec{C}, \vec{D};z) 
\nonumber \\ &&
 =   
(-1)^J 
z
\prod_{i=1}^K (\theta \!+\! A_i)
\prod_{j=1}^J (\theta \!-\! D_j \!+\! 1)
\Phi(\vec{A},\vec{B};\vec{C}, \vec{D};z) \;, 
\label{Phi:DE}
\end{eqnarray}
where $\theta = zd/dz$. 
Let us consider the case of the non-confluent function, for which the orders of
the differential equations on the l.h.s.\ and r.h.s.\ of Eq.~(\ref{Phi:DE}) are
equal to each other, viz.\
\begin{equation}
K+J = 1+ L + R \equiv p \;, 
\label{order}
\end{equation}
so that the function $\Phi$ satisfies a differential equation of order $p$.
In this case, there are $p$ linearly independent solutions of the differential
equation. 
In accordance with Takayama's algorithm \cite{theorem}, the differential
operators inverse to the operators defined by Eq.~(\ref{step-up-down}) can be
constructed, and the result of the differential reduction applied to the
function $\Phi$ has the following form: 
\begin{eqnarray}
P_0 \Phi(\vec{A}+\vec{m}_1,\vec{B}+\vec{m}_2;\vec{C}+\vec{m}_3,\vec{D}+\vec{m}_4;z) 
= 
\sum_{i=0}^{p-1} R_i \theta^i \Phi(\vec{A},\vec{B};\vec{C},\vec{D};z) 
\;, 
\label{reduction}
\end{eqnarray}
where $P_0,R_i$ are some polynomials and $\vec{m_i}$ are sets of integers.

\noindent
{\bf Theorem 1:}\\
Any Feynman diagram associated with the function $\Phi$ defined by the
Mellin-Barnes integral in Eq.~(\ref{Phi1}), under the conditions that all
parameters as well the differences between any two parameters are not integer
and Eq.~(\ref{order}) is valid, has $p$ master integrals (including all
integrals following from the original one by contracting one or more lines), 
where $p$ is defined by Eq.~(\ref{order}).
\\
This theorem follows from Takayama's analysis of the differential equations in
Eq.~(\ref{Phi:DE}) with irreducible monodromy groups. 
\\
Under the conditions of {\bf Theorem 1} and the additional condition
$C_i \neq C_j $, it is possible to close the contour of integration \cite{BD}
and to write the function $\Phi$ in terms of linear combinations of
hypergeometric functions whose series representations are well defined in the
vicinity of the point $z=0$: 
\begin{eqnarray}
\frac{z^{-r}}{C_\Phi}
\Phi(\vec{A},\vec{B};\vec{C}, \vec{D};z)
&=& 
\frac{
\prod_{i=1}^{K} \Gamma\left(A_i\right)
\prod_{j=1}^{L} \Gamma\left(C_j\right)
}
{
\prod_{k=1}^{R} \Gamma\left(B_k \right)
\prod_{l=1}^{J} \Gamma\left(D_l \right)}
{}_{p}F_{p-1}\left(\begin{array}{c|}
\vec{A}, \vec{1}-\vec{D} \\
\vec{B}, \vec{1}-\vec{C} \end{array} ~(-1)^{1+L+J} z \right) 
\nonumber \\ &&{} 
+ 
\sum_{m=1}^L z^{C_m}
\frac{
\prod_{i=1}^{K} \Gamma\left(A_i \!+\! C_m\right)
\prod_{j=1; j\neq m}^{L} \Gamma\left(C_j \!-\! C_m\right)
}
{
\prod_{k=1}^{R} \Gamma\left(B_k \!+\! C_m \right)
\prod_{l=1}^{J} \Gamma\left(D_l \!-\! C_m \right)
}
\nonumber\\
&&{}\times
{}_{p}F_{p-1}\left(\begin{array}{c|}
\vec{A} \!+\! C_m, \vec{1} \!+\! C_m \!-\! \vec{D} \\
\vec{B} \!+\! C_m, \vec{1} \!+\! C_m \!-\! \vec{\hat{C}} \end{array} ~(-1)^{L\!+\!J} z \right) 
\;,
\label{hyper}
\end{eqnarray}
where $\vec{\hat{C}}$ denotes the set of parameters $\vec{C}$ excluding $C_m$.

\noindent
{\bf Corollary 1:}\\
The number of nontrivial master integrals of a Feynman diagram satisfying the
conditions of {\bf Theorem 1} and the additional condition that $C_i \neq C_j$
is equal to the number of basic functions for any hypergeometric function on
the r.h.s.\ of Eq.~(\ref{hyper}).

\noindent
{\bf Comment 1:} \\
The application of the relation 
\begin{equation}
\Gamma(nx)  
=
n^{nx-1/2} (2\pi)^{\frac{1-n}{2}} 
\Gamma(x)
\Gamma \left( x+\frac{1}{n} \right)
\cdots 
\Gamma \left( x+\frac{n-1}{n} \right) 
\;, 
\end{equation}
where $n$ is integer and $x=t+p/q$, allows us to express any one-fold
Mellin-Barnes integral as an integral of the form defined by Eq.~(\ref{MB}).

A special consideration is necessary when some parameters or some differences
between parameters are integer.
In this case, the final expression for the differential reduction of Horn-type
hypergeometric functions has a simpler form (for details, see Ref.~\cite{BKK}).
To evaluate the dimension of the solution space of the solutions of the differential equation in
Eq.~(\ref{Phi:DE}), a classical technique can be applied \cite{Ince}.
However, for practical applications to Feynman diagrams, we only need two
particular cases, namely
case~(i) when the differences between upper ($A_i,D_j$) and lower ($B_k,C_l$)
parameters are positive integers and 
case~(ii) when some of the parameters are positive integers. 

In case~(i), the original Mellin-Barnes integral in Eq.~(\ref{Phi1}) may be
simplified as
\begin{eqnarray}
\int dt z^t
\frac{\Gamma(A+m+t)}{\Gamma(A+t)}
F(t) 
= 
(A \!+\! m \!-\!1 \!+\! \theta) (A \!+\! m \!-\! 2 \!+\! \theta) \cdots (A \!+\! \theta) 
\int dt 
z^t F(t) \;.\quad
\label{factor}
\end{eqnarray}
In this case, the order of the differential equation in Eq.~(\ref{Phi:DE}) is
$p-1$, and there are $p-1$ nontrivial master-integrals.
After the application of step-up/step-down operators to the r.h.s.\ of
Eq.~(\ref{hyper}), we obtain the hypergeometric functions $_{p-1}F_{p-2}$,
and {\bf Corollary 1} is valid.

In case~(ii), the differential equation in Eq.~(\ref{Phi:DE}) has a common
differential factor $\theta$, since 
$
z \left( \theta \!+\! 1 \right) f(z) = \theta \left( z f(z) \right) \;. 
$
In this case, the result of the differential reduction of
Eq.~(\ref{reduction}) has the following form: 
\begin{eqnarray}
P_0 \Phi(\vec{A}+\vec{m}_1,\vec{B}+\vec{m}_2;\vec{C}+\vec{m}_3,\vec{D}+\vec{m}_4;z) 
= 
\sum_{i=0}^{p-2} R_i \theta^i \Phi(\vec{A},\vec{B};\vec{C},\vec{D};z) 
\!+\! R_{p-1}(z)  
\;, 
\nonumber\\
\label{reduction:2}
\end{eqnarray}
where $P_0,R_i$ are some polynomials and $\vec{m_i}$ are sets of integers.
The same is true for the r.h.s.\ of Eq.~(\ref{hyper}), so that
{\bf Corollary 1} is again valid. 

\noindent
{\bf Corollary 2:}\\
The number of nontrivial master-integrals of the Feynman diagram associated
with the function $\Phi$ defined by Eq.~(\ref{Phi1}), under the condition that
$C_i \neq C_j$, is equal to number of basic functions for any hypergeometric
function on the r.h.s.\ of Eq.~(\ref{hyper}).

\noindent
{\bf Comment 2:}\\
The differential equation in Eq.~(\ref{Phi:DE}) can be factorized due to the
relation 
\begin{eqnarray}
z \left( \theta \!+\! 1 \!+\! a\right) f(z) = \left( \theta \!+\! a \right) \left( z f(z) \right) \;, 
\label{factorization}
\end{eqnarray}
where $a$ is an arbitrary parameter.
This equation corresponds to the factor $\Gamma(a-t) \Gamma(1-a+t)$ in the
numerator or denominator of the Mellin-Barnes integral in Eq.~(\ref{Phi1}).

\noindent
{\bf Comment 3:}\\
Also Tarasov \cite{Tarasov:1998} proposed that, in the framework of Ref.~\cite{KK},
there is a one-to-one correspondence between the number of master integrals
obtained from the IBP relations and the dimension of the solution space of a
corresponding system of differential equations.

In this way, we showed that our conjecture presented in Ref.~\cite{BKK} is
correct. 
Below, we present its multivariable generalization:

\noindent
{\bf Proposition 1:}\\
When a multivariable Mellin-Barnes integral can be presented as a linear
combination of multivariable Horn-type hypergeometric functions with rational
coefficients\footnote{%
We called such a representation hypergeometric.}
about some
points $z_i=z^0_i$, the holonomic rank of the corresponding system of linear
differential equations is equal to the holonomic rank of any hypergeometric
function in its hypergeometric representation. 

The proof of this proposition is based on the same technology, namely the
comparison of the holonomic rank of the system of differential equations in
Eq.~(\ref{MB:DE}) with the holonomic rank of each term of its hypergeometric
representation.
A two-variable example of this statement was presented in
Ref.~\cite{HYPERDIRE}.

\noindent
{\bf Conjecture 1:}\\
Any polynomial (rational) solution of a multivariable linear system of
differential equations related to a Feynman diagram can be written as a product
of one-loop bubble integrals and massless propagator or vertex integrals.
\section{Feynman diagrams}

\begin{figure}[th]
\centering
{\vbox{\epsfysize=90mm \epsfbox{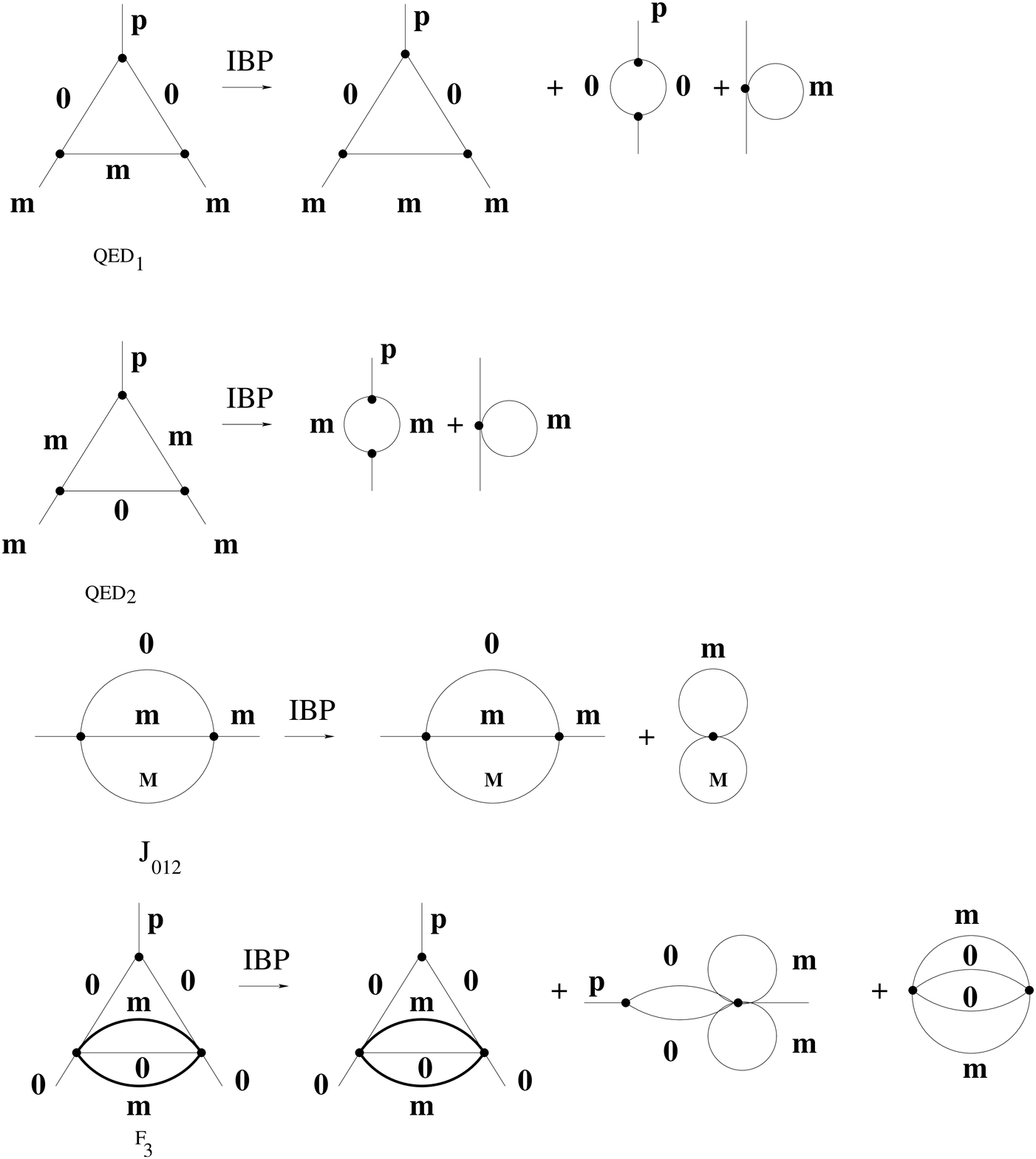}}}
\caption{
The diagrams on the r.h.s.\ emerge from the original one on the l.h.s.\ with
arbitrary positive powers of propagators by the systematic contraction of one
line. 
}
\label{vertex:qed}
\end{figure}

Several examples of Feynman diagrams corresponding to the function $\Phi$ in
Eq.~(\ref{Phi1}) were presented in Refs.~\cite{BKK,HYPERDIRE,diagrams}.
The results of the analysis performed in\break
Refs.~\cite{BKK,HYPERDIRE,diagrams}
are in agreement with {\bf Corollary 2} of the present paper. 
Nevertheless, for the illustration of the advocated technique, we consider here
the diagrams depicted in Fig.~\ref{vertex:qed}. 
In all examples, we put $r_i=0$, since non-zero values of $r_i$ do not affect
the order of the differential equations and may be easily restored by the
redefinitions $\theta_i \to \theta_i - r_i$.
\boldmath
\subsection{One-loop vertex QED$_1$}
\unboldmath

Let us consider the one-loop QED vertex diagram with one massive internal line
and two external lines being on mass shell, which is labeled QED$_1$ in
Fig.~\ref{vertex:qed}.
It is given by
\begin{eqnarray}
C_{\text{QED}_1}(\sigma_1,\sigma_2,j_3)
&\equiv&
\int
\left.
\frac{d^nk}
{
[(k-p_1)^2]^{\sigma_1}
[(k+p_2)^2]^{\sigma_2}
[k^2-m^2]^{j_3}
}
\right|_{p_1^2=p_2^2=m^2}
\label{qed1}\\
&=&
\frac{
i^{1-n} \pi^{n/2}
(-m^2)^{\tfrac{n}{2} \!-\! \sigma_{12} \!-\! j_3}
}
{
\Gamma\left( n  \!-\! \sigma_{12} \!-\!  j_3 \right)
\Gamma\left(j_3 \right)
\Gamma\left(\sigma_1 \right)
\Gamma\left(\sigma_2 \right)
}
\,\frac{1}{2 \pi i }
\int_{-i \infty}^{i \infty} dt
\left( 
-\frac{k^2}{m^2}
\right)^t 
\Gamma(-t) 
\nonumber\\
&&{}
\times
\Gamma(\sigma_1 \!+\! t)
\Gamma(\sigma_2 \!+\! t)
\Gamma\left( \sigma_{12} \!+\! j_3 \!-\! \frac{n}{2} \!+\! t \right)
\Gamma\left( n  \!-\!  j_3 \!-\! 2 \sigma_{12} \!-\! 2t \right)
\;,
\nonumber
\end{eqnarray}
where $\sigma_{12} = \sigma_1 + \sigma_2$
and $k^2=(p_1-p_2)^2$.
This diagram corresponds to a $\Phi$ function with six parameters,
\begin{equation}
C_{\text{QED}_1}(\sigma_1,\sigma_2,j_3)
= 
\Phi_{\text{QED}_1}\left(\sigma_1, \sigma_2, \sigma_{12} \!+\! j_3 \!-\! \tfrac{n}{2};-; \tfrac{n-j_3}{2} - \sigma_{12},   \tfrac{n-j_3+1}{2} - \sigma_{12}, 0;-;z \right) \;,
\end{equation}
and satisfies a third-order differential equation,
\begin{eqnarray}
&&\left[ 
\theta
(\theta \!-\! C_1) 
(\theta \!-\! C_2) 
+  
z
(\theta \!+\! A_1) 
(\theta \!+\! A_2) 
(\theta \!+\! A_3) 
\right] 
\Phi_{\text{QED}_1}\left(A_1,A_2,A_3;-;C_1,C_2,0;-;z \right)
\nonumber\\
&&= 0 \;.
\label{qed1:de}
\end{eqnarray}
With the help of differential operators,
\begin{equation}
\Phi_{\text{QED}_1}(A_i+1)  =  \left( \theta + A_i \right) \Phi_{\text{QED}_1}(A_i) \;,
\end{equation}
it is easy to reduce the value of $A_1$ (or $A_2$) to unity, 
and the one of $A_3$ to $1-C_1$ or $1-C_2$, depending on whether $j_3$ is even
or odd.
In this case, there is a double factorization of the differential equation in
Eq.~(\ref{qed1:de}) due to Eq.~(\ref{factorization}), 
\begin{eqnarray}
\theta
(\theta \!-\! C_1) 
\left[ 
(\theta \!-\! C_2) 
+  
z
(\theta \!+\! A_2) 
\right] 
\Phi_{\text{QED}_1}\left(1,A_2,1-C_1;-;C_1,C_2,0;-;z \right)  = 0 \;.
\label{qed1:de:2}
\end{eqnarray}
As a consequence, there are two polynomial solutions. 
We conclude from our analysis that there are three master integrals
corresponding to the order of the differential equation in Eq.~(\ref{qed1:de}):
one is a nontrivial function and two are polynomials. 
\boldmath
\subsection{One-loop vertex QED$_2$}
\unboldmath

Let us now consider the one-loop QED vertex diagram with two massive internal
lines and two external lines being on mass shell, which is labeled QED$_2$ in
Fig.~\ref{vertex:qed}.
It is given by
\begin{eqnarray}
C_{\text{QED}_2}(j_1,j_2,\sigma)
&\equiv&
\int
\left.
\frac{d^nk}
{
[(k-p_1)^2-m^2]^{j_1}
[(k+p_2)^2-m^2]^{j_2}
(k^2)^{\sigma}
}
\right|_{p_1^2=p_2^2=m^2}
\nonumber \\
&=&
\frac{
i^{1-n} \pi^{n/2}
(-m^2)^{\tfrac{n}{2} \!-\! j_{12} \!-\! \sigma}
\Gamma\left( n  \!-\!  j_{12} \!-\! 2 \sigma  \right)
}
{
\Gamma\left(j_1 \right)
\Gamma\left(j_2 \right)
\Gamma\left( n  \!-\!  j_{12} \!-\! \sigma  \right)
}
\,
\frac{1}{2 \pi i }
\int_{-i \infty}^{i \infty} dt
\left( 
-\frac{k^2}{m^2}
\right)^t 
\nonumber \\ &&
\times
\frac{
\Gamma(-t) 
\Gamma(j_1+t)
\Gamma(j_2+t)
\Gamma\left( j_{12} \!+\! \sigma \!-\! \frac{n}{2} \!+\! t \right)
}
{
\Gamma\left( j_{12} \!+\! 2t \right)
}
\;.
\label{qed2}
\end{eqnarray}
This diagram corresponds to a $\Phi$ function with six parameters,
\begin{equation}
C_{\text{QED}_2}(j_1,j_2,\sigma)
= 
\Phi_{\text{QED}_2}\left(j_1,j_2, j_{12} \!+\! \sigma \!-\! \tfrac{n}{2}; \tfrac{j_{12}}{2},   \tfrac{j_{12}+1}{2},0;-;-;z \right) \;.
\end{equation}
Repeatedly applying differential operators,
$
\Phi_{\text{QED}_2}(A_i+1)  =  \left( \theta + A_i \right) \Phi_{\text{QED}_2}(A_i) \;,
$
and 
$
\Phi_{\text{QED}_2}(B_i-1)  =  \left( \theta + B_i \!-\! 1 \right) \Phi_{\text{QED}_2}(B_i) \;,
$
we reduce $A_1$, $A_2$, and $B_1$ to unity, so 
that the six-parameter $\Phi$ function reduces to a four-parameter one, 
\begin{equation}
\Phi_{\text{QED}_2}\left(1,1,A; 1,B;0;-;z \right)
= 
\Phi_{\text{QED}_2}\left(1,A;B;0;-;z \right) \;,
\end{equation}
which satisfies a second-order differential equation,
\begin{eqnarray}
\theta
\left[ 
(\theta \!+\! B-1) 
+  
z
(\theta \!+\! A) 
\right] 
\Phi_{\text{QED}_2}\left(1,A;B;0;-;z \right)  = 0 \;.
\label{qed2:de}
\end{eqnarray}
Consequently, there are two master integrals, one of which is polynomial. 
This result does not depend on whether $\sigma$ is integer or not. 

\boldmath
\subsection{Two-loop sunset diagram $J_{012}$}
\unboldmath

Let us now consider the two-loop sunset diagram from Ref.~\cite{JK}, which is
given by
\begin{eqnarray}
&&
J_{012}(\sigma,\alpha,\beta)
\equiv
\int
\left.
\frac{d^n(k_1 k_2)}
{
[(k_1-p)^2]^\sigma
[k_1^2-M^2]^\alpha
[(k_1-k_2)^2-m^2]^\beta
}
\right|_{p^2=m^2}
\nonumber \\
&&=
\frac{
[i^{1-n} \pi^{n/2}]^2
(-m^2)^{n \!-\! \alpha \!-\! \sigma \!-\! \beta}
\Gamma\left( \tfrac{n}{2}  \!-\! \sigma  \right)
}
{
\Gamma\left(\sigma \right)
\Gamma\left(\alpha \right)
\Gamma(\beta)
}
\frac{1}{2 \pi i }
\int_{-i \infty}^{i \infty} dt
\left( 
\frac{M^2}{m^2}
\right)^t 
\label{j012}\\
&&{}
\times
\frac{
\Gamma\left( \sigma \!+\! \alpha \!-\! \frac{n}{2} \!+\! t \right)
\Gamma\left( \alpha \!+\! \beta \!+\! \sigma \!-\! n \!+\! t \right)
\Gamma(-t) 
\Gamma\left( \frac{n}{2} \!-\! \alpha \!-\! t \right)
\Gamma\left( 2n \!-\! \alpha \!-\! 2 \sigma \!-\! 2 \beta \!-\! 2t \right)
}
{
\Gamma\left( n \!-\! \sigma \!-\! \alpha  \!-\! t \right)
\Gamma\left( \frac{3n}{2} \!-\! \sigma \!-\! \alpha  \!-\! \beta \!-\! t \right)
}
\;.
\nonumber
\end{eqnarray}
This integral corresponds to a $\Phi$ function with eight parameters and
satisfies a fourth-order differential equation, 
\begin{eqnarray}
&& 
\theta 
\left( \theta \!-\! \frac{n}{2} \!+\! \alpha  \right)
\left( \theta \!-\! n \!+\! \tfrac{\alpha}{2} \!+\! \sigma \!+\! \beta \right)
\left( \theta \!-\! n \!+\! \tfrac{\alpha \!-\! 1}{2} \!+\! \sigma \!+\! \beta \right)
\Phi_{J_{012}}
= 
\\ && 
z
\left( 
\theta \!-\! \frac{n}{2} \!+\! \sigma \!+\! \alpha 
\right)
\left( \theta \!-\! n \!+\! \alpha \!+\! \beta \!+\! \sigma \right)
\left( \theta \!-\! n \!+\! \sigma \!+\! \alpha  \!+\! 1 \right)
\left( \theta \!-\! \frac{3n}{2} \!+\! \sigma \!+\! \alpha  \!+\! \beta \!+\! 1\right)
\Phi_{J_{012}} \;.
\nonumber 
\end{eqnarray}
With the help of step-up/step-down operators, this equation can be written as
follows: 
\begin{eqnarray}
&& 
\left( \theta \!-\! \frac{n}{2} \!+\! I_1 \right)
\left( \theta \!-\! n \!+\! I_2 \right)
\left[ 
\theta 
\left( \theta \!-\! n \!+\! \frac{1}{2} \!+\! I_3 \right)
- 
z
\left( \theta \!-\! \frac{3n}{2} \!+\! I_4 \right)
\right]
\Phi_{J_{012}} = 0 \;.
\label{J012:de}
\end{eqnarray}
Consequently, there are four master integrals, two of which are polynomial. 
All topologically possible integrals for this diagram are depicted in
Fig.~\ref{vertex:qed}.
The right-most one is polynomial in this approach. 
In order to have four master integrals, it is necessary that the diagram with
the original topology has three master integrals, which we may take to have the
propagator powers $(1,1,1)$, $(1,1,2)$, and $(1,2,1)$.
We note that this is in accordance with the results of Ref.~\cite{Tarasov}. 
However, one of these three master integrals should be polynomial.
This polynomial solution was first derived in Ref.~\cite{KK2011}.
An alternative derivation has recently be presented in Ref.~\cite{KK2012}.


\boldmath
\subsection{Three-loop vertex diagram $F$}
\unboldmath

Let us consider the vertex diagram denoted as $F_3$ in Fig.~\ref{vertex:qed},
which is given by
\begin{eqnarray}
&&
F(\vec{\sigma}_1,\vec{\sigma}_2, \vec{\sigma}_3, \alpha_1, \alpha_2)
\label{CF}\\
&& 
= 
\int
\left.
\frac{d^n(k_1k_2k_3)}
{
[(k\!-\!p_1)^2]^{\vec{\sigma_1}}
[(k\!+\!p_2)^2]^{\vec{\sigma_2}}
[k_2^2\!-\!m^2]^{\alpha_1}
[k_3^2\!-\!m^2]^{\alpha_2}
[(k_1\!+\!k_2\!+\!k_3^2)]^{\vec{\sigma_3}}
}
\right|_{p_1^2=p_2^2=0} \;,
\nonumber
\end{eqnarray}
where we have introduced a ``dressed'' massless propagator, as in Eq.~(45) of
Ref.~\cite{BKK}.
Instead of three massless lines, we consider $q_1$, $q_2$, and $q_3$ massless
lines corresponding to propagators with powers $\sigma_1$, $\sigma_2$, and
$\sigma_3$, respectively.
In the present case, this is equivalent to the redefinition
$
\sigma_j \to \sigma_j - \frac{n}{2} (q_j - 1) \;. 
$  
The Mellin-Barnes integral for the vertex diagram of Eq.~(\ref{CF}) has the
following form: 
\begin{eqnarray}
&&
F(\vec{\sigma_1},\vec{\sigma_2}, \vec{\sigma_3}, \alpha_1, \alpha_2)
\nonumber \\ &&
=
\text{Const}
\times
\frac{1}{2 \pi i }
\int_{-i \infty}^{i \infty} dt
\left( 
-\frac{p^2}{m^2}
\right)^t 
\frac{
\Gamma(\alpha_1 \!+\! t)
\Gamma(\alpha_2 \!+\! t)
\Gamma\left( \alpha_{12} \!-\! \frac{n}{2} \!+\! t \right)
\Gamma\left( \frac{n}{2} \!+\! t \right)
}
{
\Gamma\left( \alpha_{12} \!+\! 2t \right)
}
\nonumber \\ && 
\times
\frac{
\Gamma\left( \vec{\sigma_3}    \!-\! \frac{n}{2}q_3    \!-\! t \right)
\Gamma\left( \frac{n}{2}q_{13}  \!-\! \vec{\sigma}_{13}  \!+\! t \right)
\Gamma\left( \frac{n}{2}q_{23}  \!-\! \vec{\sigma}_{23}  \!+\! t \right)
\Gamma\left( \vec{\sigma}_{123} \!-\! \frac{n}{2}(q_{123} \!-\! 1) \!-\! t \right)
}
{
\Gamma\left( \frac{n}{2}(q_3 \!+\! 1) \!-\! \sigma_3    \!+\! t \right)
\Gamma\left( \frac{n}{2}q_{123}  \!-\! \vec{\sigma}_{123}  \!+\! t \right)
}
\;.\qquad
\label{F:MB}
\end{eqnarray}
where $\text{Const}$ is some constant irrelevant for the differential equation.

This diagram corresponds to a $\Phi$ function with twelve parameters. 
The first simplification arises from the step
\begin{equation}
\frac{
\Gamma(\alpha_1 \!+\! t)
\Gamma(\alpha_2 \!+\! t)
}
{
\Gamma\left( \alpha_{12} \!+\! 2t \right)
}
\to 
\frac{
\Gamma(I_1 \!+\! t)
}
{
\Gamma\left( \frac{1}{2} \!+\!  I_{2} \!+\! t \right)
} \;,
\end{equation}
where $I_a$ are integers. 
The further procedure strongly depends on whether the values of $q_1$ and $q_2$
are even or odd.
For simplicity, we put $q_1=q_2=q_3=1$ and denote this integral as $F_3$.
Then the following step is valid:
\begin{equation}
\left.
\frac
{
\Gamma\left( \frac{n}{2}q_{13}  \!-\! \vec{\sigma}_{13}  \!+\! t \right)
\Gamma\left( \frac{n}{2}q_{23}  \!-\! \vec{\sigma}_{23}  \!+\! t \right)
}
{
\Gamma\left( \frac{n}{2}(q_3 \!+\! 1) \!-\! \sigma_3    \!+\! t \right)
}
\right|_{q_1=q_2=q_3=1}
\to 
\Gamma\left( n \!+\! I_3  \!+\! t \right)
\;,
\end{equation}
so that $F_3$ satisfies a homogeneous differential equation of fourth order
that may be written in the following form: 
\begin{eqnarray}
&& 
\left(
\theta \!+\! \frac{1}{2} \!+\! I_1
\right)
\left(
\theta \!+\! \frac{3}{2}n \!+\! I_2
\right)
\left(
\theta \!+\! \frac{n}{2} \!+\! I_3
\right)
\left(
\theta \!+\! n \!+\! I_4
\right)
\Phi_{F_3} 
\nonumber \\ &&
= 
z
\left(
\theta \!+\! I_5
\right)
\left(
\theta \!-\! \frac{n}{2} \!+\! I_7
\right)
\left(
\theta \!+\! \frac{n}{2} \!+\! I_6
\right)
\left(
\theta \!+\! n \!+\! I_8
\right)
\Phi_{F_3} 
\;.
\label{F:de}
\end{eqnarray}
Applying step-up/step-down operators, this equation can be converted to the
form
\begin{eqnarray}
&& 
\left(
\theta \!+\! \frac{n}{2} 
\right)
\left(
\theta \!+\! n 
\right)
\left[
\left(
\theta \!+\! \frac{1}{2} \!+\! I_1
\right)
\left(
\theta \!+\! \frac{3}{2}n \!+\! I_2
\right)
-
z
\left(
\theta \!+\! I_3
\right)
\left(
\theta \!-\! \frac{n}{2} \!+\! I_4
\right)
\right]
\Phi_{F_3} 
= 0 
\;.\qquad
\label{F:de:2}
\end{eqnarray}
Consequently, for the vertex diagram $F_3$ there are four master integrals,
two of which are polynomial. 
All topologically possible integrals for this case are shown in
Fig. \ref{vertex:qed}.
In order to have four master integrals, it is necessary that the diagram with
the original topology has two master integrals.

\section{Discussion and conclusions}

We proposed a novel way of finding linear systems of homogeneous differential
equations for Feynman diagrams with arbitrary powers of propagators.
It is based on the Mellin-Barnes representation and does not rely on the IBP
relations \cite{ibp}.
Systems of equations such as Eq.~(\ref{MB:DE}) are the left ideals in the Weyl
algebra of linear differential operators with polynomial coefficients. 
Exploiting the Gr\"obner basis technique \cite{Grobner}, the original diagrams
may be explicitly reduced to sets of basis functions, and their holonomic
ranks, i.e.\ the numbers of their linearly independent solutions, may be
evaluated.

For the one-variable case, we presented a very simple algorithm for counting
polynomial (rational) solutions of differential equations arising from the
Mellin-Barnes representations associated with Feynman diagrams.
This reduces the problem to the one of factorizing the homogeneous differential
equation over the field of polynomials.
We presented a few simple examples to illustrate our technique. 
The generalization of this algorithm to the multivariable case is nontrivial. 

With the help of the new technology presented here, we proved the conjecture
formulated in Ref.~\cite{BKK} regarding the counting of the numbers of master
integrals via hypergeometric representations.
This result may be useful for searching polynomial (rational) solutions of
multivariable Feynman diagrams. 

We suggest that any polynomial (rational) solution corresponds to a product of
one-loop bubbles and massless single-scale diagrams with coefficients that are
products of Gamma functions (see {\bf Conjecture 1}).
Even in the one-variable case, such a correspondence does not emerge from the
application of standard IBP relations, as was pointed out in
Refs.~\cite{KK2011,KK2012} for the case of the two-loop sunset diagram
$J_{012}$.
With help of the technique presented here, all such algebraic relations between
master integrals of the type studied in Ref.~\cite{BKK,HYPERDIRE,diagrams} may
be easily derived.

Hypergeometric functions provide us with a remarkable tool for deepening our
understanding of the mathematical structures underlying Feynman diagrams,
and the present analysis allows us to draw the following picture. 
From the fact that any Feynman diagram with arbitrary powers of propagators is
reducible to a set of master integrals, including bubble diagrams and massless
propagators, it follows that a given Feynman diagram corresponds to a special
function with a reducible monodromy group (see also
Ref.~\cite{monodromy,recent}). 
The dimension of the irreducible part of the monodromy group, which is equal to
the dimension of the solution space of the Pfaff system of differential
equations, is equal to the number of master integrals generated via IBP
relations, provided the full set of the latter is exploited.
It is interesting to note that the simplest way of avoiding the reducibility of
the monodromy group is to introduce different non-integer parameters for each
propagator.
This may be regarded as a generalization of the analytical regularization.

The proposed method to derive differential equations from the Mellin-Barnes
representations of Feynman diagrams with non-unit values of $z_i$ can be
directly applied to study massless propagator diagrams at higher loop orders.
In this case, we have to tackle with multiple (threefold and higher)
Mellin-Barnes integrals.
We shall return to this issue in a future publication.

\noindent 
{\bf Acknowledgments} \\
We are grateful to Oleg Veretin for useful discussions and to Oleg Tarasov for
his interest to our work.
This work was supported in part by the German Federal Ministry for Education
and Research BMBF through Grant No.\ 05~HT6GUA, by the German Research
Foundation DFG through the Collaborative Research Centre No.~676
{\it Particles, Strings and the Early Universe---The Structure of Matter and
Space-Time}, and by the Helmholtz Association HGF through the Helmholtz
Alliance Ha~101 {\it Physics at the Terascale}.


\begin{thebibliography}{99}

\bibitem{Regge}
T.~Regge,
Algebraic Topology Methods in the Theory of Feynman Relativistic Amplitudes,
Battelle Rencontres: 1967 Lectures in Mathematics and Physics,
C.M. DeWitt, J.A. Wheeler (Eds.),
(W.A. Benjamin, New York, 1968), pp.~433--458.

\bibitem{dimreg}
G.~'t~Hooft, M.~Veltman,
Nucl.\ Phys.\ B 44 (1972) 189.

\bibitem{DE}
A.V.~Kotikov,
Phys.\ Lett.\ B 254 (1991) 158;\\
E.~Remiddi,
Nuovo Cim.\ A 110 (1997) 1435.

\bibitem{ibp}
F.V.~Tkachov, 
Phys.\ Lett.\ B 100 (1981) 65;\\
K.G.~Chetyrkin, F.V.~Tkachov, 
Nucl.\ Phys.\ B 192 (1981) 159.

\bibitem{review}
M.~Argeri, P.~Mastrolia,
Int.\ J.\ Mod.\ Phys.\  A 22 (2007) 4375,
arXiv:0707.4037 [hep-ph].

\bibitem{idea}
M.Yu.~Kalmykov, V.V.~Bytev, B.A.~Kniehl, B.F.L.~Ward, S.A.~Yost,
PoS (ACAT08) 125,
arXiv:0901.4716 [hep-th].

\bibitem{calan}
C.~de Calan, A.P.C.~Malbouisson,
Annales Poincare Phys.\ Theor.\ 32 (1980) 91;\\
C.~de Calan, F.~David, V.~Rivasseau,
Commun.\ Math.\ Phys.\ 78 (1981) 531;\\
C.~de Calan, A.P.C.~Malbouisson,
Commun.\ Math.\ Phys.\ 90 (1983) 413.

\bibitem{BD}
E.E.~Boos, A.I.~Davydychev,
Theor.\ Math.\ Phys.\  89 (1991) 1052
[Teor.\ Mat.\ Fiz.\ 89 (1991) 56].

\bibitem{smirnov}
V.A.~Smirnov,
Evaluating Feynman integrals,
Springer Tracts Mod.\ Phys.\ 211 (2004) 1;\\ 
V.A.~Smirnov,
Feynman integral calculus
(Springer-Verlag, Berlin, 2006).

\bibitem{Grobner}
M.~Saito, B.~Sturmfels, N.~Takayama, 
Gr\"obner Deformations of Hypergeometric Differential Equations
(Springer-Verlag, Berlin, 2000).

\bibitem{cartan}
E.~Cartan, 
Les Syst\`emes Differentialles Ext\'eriers et Leurs Applications
G\'eom\'etriques
(Herman, Paris, 1945).

\bibitem{theorem}
N.~Takayama,
Japan J.\ Appl.\ Math.\ 6 (1989) 147. 

\bibitem{KK}
M.~Kashiwara, T.~Kawai,
Publ.\ Res.\ Inst.\ Math.\ Sci.\ Kyoto 12 (1977) 131;\\
M.~Kashiwara, T.~Kawai,
Commun.\ Math.\ Phys.\ 54 (1977) 121;\\
T.~Kawai, H.P.~Stapp,
Commun.\ Math.\ Phys.\ 83 (1982) 213.

\bibitem{Speer}
E.R.~Speer
J.\ Math.\ Phys.\ 9 (1968) 1404.

\bibitem{Tarasov:1998}
O.V.~Tarasov,
Acta Phys.\ Polon.\ B 29 (1998) 2655,
arXiv:hep-ph/9812250.

\bibitem{D}
A.V.~Smirnov, A.V.~Petukhov,
Lett.\ Math.\ Phys.\ 97 (2011) 37,
arXiv:1004.4199 [hep-th].

\bibitem{Blumlein}
J.~Bl\"umlein, S.~Klein, C.~Schneider, F.~Stan,
J. Symbol.\ Comput.\ 47 (2012) 1267,
arXiv:1011.2656 [cs.SC].

\bibitem{beukers}
F.~Beukers,
arXiv:1101.0493 [math.AG].

\bibitem{gelfand}
I.M.~Gelfand, M.M.~Kapranov, A.V.~Zelevinsky,
Funck.\ Anal.\ i Priloz.\ 23 (1989) 94;\\
I.M.~Gelfand, M.M.~Kapranov, A.V.~Zelevinsky,
Adv.\ Math.\  84 (1990) 255;\\
I.M.~Gel'fand, M.I.~Graev, V.S.~Retakh,
Russian Math.\ Surveys 47 (1992) 1.

\bibitem{resolution}
V.A.~Smirnov,
Phys.\ Lett.\  B 460 (1999) 397,
arXiv:hep-ph/9905323; \\
J.B.~Tausk,
Phys.\ Lett.\  B  469 (1999) 225,
arXiv:hep-ph/9909506;\\
C.~Anastasiou, A.~Daleo,
J. High Energy Phys.\ 0610 (2006) 031,
arXiv:hep-ph/0511176;\\
M.~Czakon,
Comput.\ Phys.\ Commun.\ 175 (2006) 559,
arXiv:hep-ph/0511200;\\
A.V.~Smirnov, V.A.~Smirnov,
Eur.\ Phys.\ J.\  C 62 (2009) 445,
arXiv:0901.0386 [hep-ph].

\bibitem{BKK}
V.V.~Bytev, M.Yu.~Kalmykov, B.A.~Kniehl,
Nucl.\ Phys.\ B 836 (2010) 129,
arXiv:0904.0214 [hep-th].

\bibitem{Ince}
E.L.~Ince,
Ordinary Differential Equations 
(Dover Publications, New York, 1956).

\bibitem{HYPERDIRE}
V.V.~Bytev, M.Yu.~Kalmykov, B.A.~Kniehl,
arXiv:1105.3565 [math-ph].

\bibitem{diagrams}
M.Yu.~Kalmykov,
J.\ High\ Energy\ Phys.\ 0604 (2006) 056,
arXiv:hep-th/0602028;\\
V.V.~Bytev, M.~Kalmykov, B.A.~Kniehl, B.F.L.~Ward, S.A.~Yost,
arXiv:0902.1352 [hep-th];\\
S.A.~Yost, V.V.~Bytev, M.Yu.~Kalmykov, B.A.~Kniehl, B.F.L.~Ward,
arXiv:1110.0210 [math-ph].

\bibitem{JK}
F.~Jegerlehner, M.Yu.~Kalmykov,
Nucl.\ Phys.\ B 676 (2004) 365,
arXiv:hep-ph/0308216.

\bibitem{Tarasov}
O.V.~Tarasov,
Nucl.\ Phys.\  B 502 (1997) 455,
arXiv:hep-ph/9703319.

\bibitem{KK2011}
M.Yu.~Kalmykov, B.A.~Kniehl,
Phys.\ Lett.\ B 702 (2011) 268,
arXiv:1105.5319 [math-ph].

\bibitem{KK2012}
  B.A.~Kniehl, A.V.~Kotikov,
  Phys.\ Lett.\ B 712 (2012) 233,
  arXiv:1202.2242 [hep-ph].

\bibitem{monodromy}
G.~Ponzano, T.~Regge, E.R.~Speer, M.J.~Westwater,
Commun.\ Math.\ Phys.\ 15 (1969) 83;\\
G.~Ponzano, T.~Regge, E.R.~Speer, M.J.~Westwater,
Commun.\ Math.\ Phys.\ 18 (1970) 1;\\
T.~Regge, E.R.~Speer, M.J.~Westwater,
Fortsch.\ Phys.\ 20 (1972) 365.

\bibitem{recent}
A.~Connes, M.~Marcolli,
J.\ Geom.\ Phys.\ 56 (2006) 55,
arXiv:hep-th/0504085;\\
S.~Bloch, H.~Esnault, D.~Kreimer,
Commun.\ Math.\ Phys.\ 267 (2006) 181,
arXiv:math/0510011 [math-ag];\\
M.~Marcolli,
arXiv:0804.4824 [math-ph].

\end{thebibliography}
\end{document}